\documentclass[final]{aipproc}
\layoutstyle{6x9}
\begin{document}

\title{Magnetic Fields in High-Density Stellar Matter}
\classification{97.60.Gb; 95.30.Qd; 97.10.Ld; 97.60.Jd}
\keywords{pulsars - neutron stars - equations of state - magnetic
fields - magnetohydrodynamics - superfluidity - superconductivity
- gamma ray bursts - quark matter - combustion processes.}

\author{Germ\'an Lugones}{address={Dipartimento di Fisica
``Enrico Fermi'' Universit\`a di Pisa and INFN Sezione di Pisa,\\
largo B. Pontecorvo 3, I-56127 Pisa, Italy},
email={lugones@df.unipi.it}    }

\begin{abstract}
I briefly review some aspects of the effect of magnetic fields in
the high density regime relevant to neutron stars,  focusing
mainly on compact star structure and composition,
superconductivity, combustion processes, and gamma ray bursts.
\end{abstract}

\maketitle

\section{The ``Magnetic Field'' of ``Neutron stars''}

A large body of evidence now identifies the presence of strong
magnetic fields at (or very near) the surface of neutron stars.
Except for a few cases, all the estimations of the magnetic fields
of these objects come from the timing measurements of pulsar
slow-down. The pulsar is usually modelled as a rapidly rotating
neutron star with a dipolar magnetic field configuration in which
the rotation axis is not aligned with the dipole axis. Since the
period of the pulsed emission is associated with the rotation
period $P$, accurate observations of $P$ and its time derivatives
$\dot{P}$, $\ddot{P}$, and even the third derivative of $P$, have
been used to shed light on pulsar dynamics. Within this framework
the dynamic equation reads $\dot{\Omega} = K \Omega^3$, with $K
\equiv (R^6  B_p^2 \sin^2\alpha)/(6 c^3 I)$, and $\Omega = 2 \pi
/P$. Therefore, some properties of pulsars are plausible of
interpretation in terms of timing data, e.g.:

\begin{itemize}
\item the component of the $B$-field  \textit{at} the magnetic
pole \textit{along} the rotation axis
\begin{equation} B_{\perp} =  B_p \sin\alpha = \sqrt{\frac{6 c^3
I}{R^6} \frac{P \dot{P} }{(2\pi)^2}}
\end{equation}
\item the pulsar age t (or alternatively the rotational period at
birth $P_i$),
\begin{equation} t = \tau \bigg( 1- \frac{P_i^2}{P^2} \bigg)
\label{t}
\end{equation}
\end{itemize}

\noindent where $\tau \equiv - P / 2 \dot{P}$ is the so called
characteristic age, $I$ is the moment of inertia of the neutron
star, $R$ its radius, and $\alpha$ is the angle between the
rotation axis and the dipole axis. Note that, for determining
$B_{\perp}$ the values of $I$ and $R$ must be assumed based mostly
on theoretical arguments.

A test of the reliability of these estimates, can be obtained from
an independent determination of the age of the pulsar, which is
possible for pulsars associated with supernova remnants (SNRs).
Assuming that a given pulsar and a given SNR were born together,
we can identify $t_{SNR}$ (the SNR age) with the pulsar age $t$ in
Eq. (\ref{t}). Since within the dipole model $\tau$ is an upper
limit for $t$, a comparison of $t_{SNR}$ with $\tau$ provides a
relevant test for pulsar dynamic models. Another interesting
feature is that also $P_i$ can be inferred from timing
observations through Eq. (\ref{t}). For some associations it is
found $t_{SNR}  \leq \tau$. The straightforward interpretation in
these cases  is that the pulsar was born with a small $P_i$
(probably close to the mass shed limit $P_{break}$) and therefore
$t \leq \tau$. The cases for which $t_{SNR} \ll \tau$ are usually
understood by assuming that the pulsar was born with $P_i$ very
close to the present value of $P$. The associations with $t_{SNR}
> \tau$ cannot be explained within the dipolar model. This is the
case of, e.g.,

\begin{itemize}
\item PSR J1846-0258 with a characteristic age $\tau = 700$ yrs
while the associated SNR Kes75 has an inferred age $t_{SNR}
> 2000$ yrs \cite{snr1},
\item the association PSR B1757-24 / SNR G5.4-1.2 \cite{snr2}.
\end{itemize}

\noindent These discrepancies are usually attributed to different
causes:

\begin{itemize}
\item the association is false, e.g. the pulsar could be a
foreground object unrelated to the SNR, or the estimated SNR age
is no reliable (see \cite{snr2}),

\item the pulsar spin-down torque is not due to pure magnetic
dipole radiation, but other physical agents contribute to the spin
down.
\end{itemize}

Another test is possible if $B$ is known independently. A case is
1E1207.4-5209 \cite{bignami2003} for which $B \sim 8 \times
10^{10}$ G has been determined from the observation of cyclotron
lines (in disagreement with the dipole model that predicts
$B_{\perp} = 2-3 \times 10^{12}$ G). Note that the characteristic
age of 1E1207.4-5209 is $\tau \approx 4.5 \times 10^5$ yrs  while
the age of the associated SNR PKS 1209-51/52 is only $t_{SNR} <
10^4$ yrs \cite{bignami2003,pavlov2002}. Then, from the point of
view of the age determination the association does not appear as
contradictory with the dipole model since the age can be adjusted
assuming $P_i \sim P_0$. This stresses the obvious fact that $t <
\tau$ is a necessary condition to believe in the dipole model, but
not a sufficient one.

The \textit{braking index} defined by $n \equiv \Omega
\ddot{\Omega} / \dot{\Omega}^2$ is the  quantity most usually
employed to study the pulsar braking behavior to second order.
Within the dipole model $n$ should be equal to 3 at any time.
Accurate determinations of $n_{obs}$ have been reported in the
literature only for five pulsars. The observed values are all
smaller than 3, i.e. in contradiction with the dipole model. A
number of scenarios have been proposed  to explain the
discrepancies. In general, they are attributed to a time variation
of $K \equiv (R^6  B_p^2 \sin^2\alpha)/(6 c^3 I)$ associated with:
a decoupling of a portion of the star's liquid interior from the
external torque \cite{Link-Epstein}, or the time evolution of the
moment of inertia $I$ due to phase transitions in the core of a
rotating neutron star \cite{Weber-Glendenning}, variation in the
angle $\alpha$ between the magnetic and rotation axes
\cite{AllenHorvath}, or magnetic field decay \cite{Bdecay}. There
are also many pulsars for which large (and in some cases negative)
measured values of $\ddot{\Omega}$ and $n_{obs}$ have been
obtained. These disagreements are thought to be caused by timing
noise and the recovery from glitches \cite{Lyne1998}. Although a
definitive explanation for the disagreement between theory and
observations is still lacking, it seems clear that other physical
agents than pure dipolar emission govern the pulsar dynamics (at
least) to second order.

Another interesting fact is that there are a few radio pulsars
with \emph{inferred} $B$-fields above the quantum critical field
$B_{ce} = m_e^2 c^3 / e \hbar \simeq 4.41 \times 10^{13}$ G, above
which some models \cite{baring} predict that radio emission should
not occur \cite{mclaughlin,camilo}. This group of ``high-field''
radio-pulsars may actually constitute a larger fraction of the
total pulsar population since there are selection effects that
make difficult the detection of long period radio pulsars. This
represents a challenge for pulsar emission models (see Melrose in
this volume, and references therein). Alternatively, the high
inferred $B$-fields could be simply the result of the very
schematic picture offered by a purely dipole model. In this sense,
other dynamic descriptions (e.g., pulsar disks, propellers
\cite{propeller}) may gain new interest.

\section{Effects of $B$ on neutron Star Structure and composition}

Since the fields in the interior of neutron stars can be even larger
than those believed to exist at the surface, it is necessary to
explore the properties of bulk matter in the presence of large and
also huge magnetic fields. An upper limit to the neutron star
magnetic-field strength follows from the virial theorem of
magneto-hydrostatic equilibrium: $2T + 3 \Pi + W + \mathcal{M}$.
Since $2T + 3 \Pi$ is always positive, the magnetic energy of the
neutron star can never exceed its gravitational binding energy. For
a star of mass $M$ and radius $R$ this gives $(4/3 \pi R^3)(B^2/8
\pi) \sim 3 G M^2/5R$, or

\begin{equation}
B_{max} = 1.4 \times 10^{18} \bigg( \frac{M}{M_{\odot}}  \bigg)
\bigg( \frac{R}{10 \mathrm{km} }  \bigg)^{-2} \mathrm{G}.
\end{equation}

Concerning the equation of state, the essential distinction
between zero and non-zero magnetic fields comes from the different
way of counting the energy states of the particles due to the
modification of the phase space introduced by $B$. Magnetic fields
quantize the orbital motion of the charged particles into Landau
levels. For a non-relativistic free particle of charge $q$ and
mass $m$ in a constant magnetic field (along the $z$ axis), the
kinetic energy of the transverse motion becomes $E_{\perp}
\rightarrow (n_L + 1/2) \hbar \omega_c $ with $n_L = 0,1,2$...,
where $\omega_c = |q| B / (m c)$ is the cyclotron frequency. The
transverse motion of the electrons becomes relativistic when
$\hbar \omega_c  > m_e c^2$, i.e. for magnetic fields larger that
the critical value:
\begin{equation}
B_{ce} = \frac{m_e^2 c^3}{q_e \hbar} = 4.41 \times 10^{13}
\mathrm{G}.
\end{equation}
\noindent For such large fields the free-particle energy has the
following relativistic expression:
\begin{equation}
E = [c^2 p_z^2 + m^2 c^4 (1 + 2 n_L \beta) ]^{1/2}
\end{equation}
\noindent with $\beta = B / B_{ce}$  \cite{Johnson1949}.

At high densities, the magnetic-field effect on the bulk equation
of state is expected to become negligible as more and more Landau
levels are filled. However, where the density is not too high, the
electrons reside mainly in low-lying Landau levels and important
modifications are expected with respect to the zero-field case.
The critical density $\rho_B$ below which the effect of Landau
quantization is important can be roughly estimated by considering
a degenerate electron gas in a constant magnetic field. The
electron Fermi momentum $p_F$ is obtained from $n_e = (e B/hc) (2
p_F/h)$. The electrons occupy only the ground Landau level when
the Fermi energy $p_F^2 / (2 m_e)$ is less than the cyclotron
energy $\hbar \omega_{ce}$. From this we find $\rho_B = 7.09
\times 10^3 Y_e^{-1}  B_{12}^{3/2} \mathrm{g cm^{-3}}$, being $Y_e
= Z/A$ the number of electrons per baryon. In general, we can use
the condition $\rho_B > \rho$, or $B_{12} > 27 (Y_e
\rho_6)^{2/3}$, to estimate the critical value of $B$ above which
Landau quantization will affect physics at a density $\rho$.

From this simple analysis we see that in neutron stars, magnetic
fields are important for understanding the behavior of the surface
layers. The surface layers are expected to be composed of
$^{56}$Fe (residual from the supernova explosion that originated
the neutron star), and also of lighter elements that can fall onto
the NS surface from the supernova remnant, from the surrounding
interstellar medium, or from a binary companion. This, together
with eventual nuclear reactions and weak interactions at the
surface during accretion, may result in a quite complex
composition.  On the other hand, strong gravity separates the
accreted elements \cite{Illarionov1980}. The lightest ones go up,
and a hydrogen-helium envelope is likely to form on the top of the
surface unless it has completely burnt out.  The superficial
layers can be in a gaseous state (atmosphere) or a condensed state
(liquid or solid surface), depending on surface temperature,
magnetic field and chemical composition \cite{Zavlin2002}. For a
hydrogen-helium envelope, a condensed surface forms at low
temperatures ($T < 10^6$ K) and very strong magnetic fields ($B >
10^{15}$ G) while for iron it depends on the cohesive properties
of the iron condensate \cite{Lai2001}.

On the other hand, it has long been recognized that neutron stars
are sources of soft X rays during the $\sim 10^5 -  10^6$ years of
the cooling phase after their birth in supernova explosions.  A
systematic study of X-ray emission from isolated neutron stars has
shown that in general there are two different components of the
X-ray emission, thermal and non-thermal. The non-thermal component
with a power-law spectrum, is believed to originate from the
pulsar's magnetosphere, while the thermal component is emitted
from the surface layers of the neutron star. Magnetic fields
affect dramatically the emergent spectra of the thermal emission
reflecting the changes in the atomic structure, and may also
produce a strong polarization of the emission. Also, the geometry
of the magnetic field may introduce an angular dependence of
opacities and equations of state that lead to an anisotropic
cooling and consequently a strong anisotropy of the thermal
emission \cite{Pavlov1978,Shibanov1996}. An accurate knowledge of
the modifications introduced by the magnetic field in the thermal
component is particularly interesting because this radiation
contains important information about neutron star properties such
as its surface temperature, magnetic field, and chemical
composition, as well as the NS radius and mass, that are
ultimately related to the internal composition (see
\cite{Zavlin2002} for a detailed analysis).


The inelastic scattering off electrons which are quantized in the
strong magnetic fields of neutron stars produce cyclotron lines in
the spectra providing a tool for direct measurements of neutron
star magnetic fields. Cyclotron resonances have been seen in a
handful of neutron stars in binary systems, the first one being
Hercules X-1. Only recently the first direct measure of an
isolated neutron star magnetic field has been possible
\cite{bignami2003}: A long X-ray observation of 1E1207.4-5209 with
XMM-Newton has shown three distinct features, regularly spaced at
0.7, 1.4 and 2.1 keV which are interpreted as cyclotron absorption
from electrons, resulting a magnetic field strength of $8 \times
10^{10}$ G \cite{bignami2003}. As mentioned in the Introduction,
this field strength differs by two orders of magnitude from the
value obtained from timing measurements of the pulsar slowdown,
challenging the most accepted models of pulsar dynamic behavior.


It is interesting to check whether magnetic fields are relevant
near the center of the neutron star core since huge field
strengths may be expected in this region. Near the center, the
density is well above the nuclear saturation density ($\rho_0 =
2.7 \times 10^{14} \mathrm{g ~ cm^{-3}}$) and matter may be
assumed to be composed mainly by neutrons and protons, with a
substantial amount of other charged and uncharged baryons, such as
the hyperons $\Lambda^0$, $\Sigma^{+}$, $\Sigma^{0}$,
$\Sigma^{-}$, $\Xi^{-}$, and $\Xi^{0}$ (but see below). These
baryons are in $\beta$-equilibrium with negatively charged leptons
(electrons and muons). In this context, Landau quantization leads
to a softening of the EOS relative to the case in which magnetic
fields are absent, and increases proton and lepton abundances
(above $5 \times 10^{18}$ G). Besides Landau quantization, it is
also important the effect of polarizing the spin (or magnetic
moments) of the nucleons and hadrons. This produces an overall
stiffening of the equation of state and a further suppression of
hyperons. In strong magnetic fields, contributions from the
anomalous magnetic moments of nucleons and hadrons must also be
considered. All these effects can be incorporated in a
relativistic mean-field equation of state including not only
nucleons but also hyperons \cite{Prakash}. General relativistic
magneto-hydrostatic calculations with this equations of state show
that maximum average fields within a stable neutron star are
limited to values $B \sim 1 - 3 \times 10^{18} G$, in agreement
with the simple estimates given above. Predictably, this field
strength is not large enough to influence particle compositions or
the matter pressure considerably.

\section{Superfluidity and Superconductivity}

There are two qualitatively different ideas about the magnetic
field inside neutron stars, namely: a)  it is present in the
entire star (core and crust), b) it is located mainly in the
crust. This depends strongly on the mechanism responsible for the
generation of $B$ (see Reisseneger in this Volume) and on the
appearance of superconducting phases in neutron star matter (this
Section).

Since neutron star interiors behave in most cases as highly
degenerate Fermi systems,  ($k_B T \ll \mu$, being $\mu$ the
chemical potential of neutron star matter), the presence of any
attractive channel in the particle interactions is expected to
originate a superfluid (and a superconductor in the case of
charged particles). There are two different classes of
superconductors, which affect the magnetic field differently. Type
I superconductors exhibit a very sharp transition to a
superconducting state and perfect diamagnetism, the ability of
expelling an already present magnetic field from the
superconductor when it is cooled below the critical temperature
(the so called Meissner effect). Type II superconductors  differ
from type I in that their transition from a normal to a
superconducting state is gradual across a region of partially
superconducting behavior (characterized by two critical fields
$H_{c1}$ and $H_{c2}$). For magnetic fields smaller than $H_{c1}$
there exists a Meissner effect. For $H_{c1} < H < H_{c2}$, type II
superconductors show a partially superconducting behavior. In this
range of $H$ there exist some rather novel ``mesoscopic''
phenomena like flux-lattice vortices, widely observed and studied
in terrestrial fluids. Flux-lattice vortices are tubes of
electrical current induced by an external magnetic field where
superconductivity is suppressed. For $H > H_{c2}$ the vortices are
too dense to maintain the condensate, and the normal phase is once
again restored.

From a theoretical point of view, the criterion that determines
whether a superconductor is of type I or type II is the
Ginzburg-Landau parameter $\kappa$, obtained from the
Ginzburg-Landau equations \cite{Ginzburg1950}. In their original
paper, Ginzburg and Landau \cite{Ginzburg1950} analyzed the energy
of the interface between a normal and a superconducting phase,
kept in equilibrium in the bulk by an external magnetic field at
the critical value. They found analytically that the surface
energy vanishes at $\kappa = \kappa_c  = 1/\sqrt{2}$. The physical
meaning of this critical value was clarified further by Abrikosov
\cite{Abrikosov1957}: it represents the demarcation line between
type I ($\kappa < 1/\sqrt{2}$) and type II ($\kappa > 1/\sqrt{2}$)
superconductors.

In neutron stars, several types of baryon pairing are believed to
appear due to the attractive component of the nuclear force which
allows binding into Cooper pairs through the well-known BCS
mechanism. In the inner crust region the $^1S_0$ partial wave of
the nucleon-nucleon interaction is attractive and neutrons will
form $^1S_0$ pairs. This happens in the range $10^{-3} \rho_0 <
\rho < 0.7 \rho_0$, where $\rho_0 = 2.7 \times 10^{14} \mathrm{g ~
cm^{-3}}$ is the saturation density of symmetric nuclear matter
\cite{Tanigawa2004}. In the core region  having $\rho > 0.7 \rho_0
$, the $^3P_2$ partial wave of the nucleon-nucleon interaction
becomes attractive enough leading to $^3P_2$ neutron pairing. In
contrast, the $^1S_0$ partial wave would become repulsive and the
neutrons would cease to pair in the $^1S_0$ state. Instead, the
$^1S_0$ proton pairs are predicted to appear owing to its fraction
smaller than neutrons. At much higher baryon density $\rho > 2
\rho_0$ various hyperons may emerge; some kinds of them possibly
form pairs in the same way as nucleons do.

Within a conventional picture, the proton superconductor is
believed to be type II  supporting a lattice of magnetic flux
tubes that carry the magnetic flux through the neutron star. For
some values of the proton correlation length and the London
penetration depth, the distant proton vortices repel each other
leading to formation of a stable vortex lattice. In addition, the
rotation of the neutron star causes a lattice of quantized
vortices to form in the superfluid neutron state, similar to the
observed vortices that form when superfluid Helium is rotated.
However, according to recent work \cite{Zhitnitsky2005},  a
detailed analysis of both the proton and neutron Cooper pair
condensates, indicates that the superconductor may in fact be type
I with critical temperature $H_c \sim 10^{14}$ G and would exhibit
the Meissner effect. This is in agreement with astrophysically
based arguments  indicating that the conventional picture of a
neutron star as a type-II superconductor may have to be
reconsidered \cite{Link2003}.

If the central densities of compact stars exceed the density of
deconfinement phase transition to a quark matter phase, the
deconfined quark matter can settle in a color superconducting
phase if there exist any attractive channel of the quark-quark
interaction. This subject was already addressed in the early 1980s
\cite{Bailin1984} but came back a few years ago since the
realization that the typical superconducting gaps in quark matter
may be much larger than those predicted in the early works
($\Delta$ as high as $\sim 100$ MeV) \cite{newgap}. There exist
many phases of color superconducting quark matter. Whereas at the
lower densities (near $\rho_0$) the phase structure is thought to
be very rich, it seems clear that at very large densities the
lowest energy state is the so called Color-Flavor-Locked (CFL)
phase \cite{Alford1999} where  quarks of all flavors and colors
pair. The effect of the pairing gap on the equation of state and
on the structure of compact stars  may be extremely important at
densities near the zero pressure point (i.e. near the surface of
the star). Estimations show that there is much room for paired
quark matter being more stable than nuclear matter and atomic
nuclei \cite{Lugones2002prd}, implying that there may exist
``neutron stars'' formed by CFL strange matter from the center up
to the surface \cite{Lugones2003aa}.

According to early calculations \cite{Bailin1984}, the quark
superconductor was though to be marginally type I with a
zero-temperature critical field $H_c \sim 10^{16}$ G. This picture
has been recently reconsidered
\cite{Alford2000,Iida2002,Blashke2002,Giannakis2003,Iida2004}.
Quark matter would make the gluons massive (there is a color
Meissner effect) but allows a ``modified'' photon (a combination
of electromagnetic and some gluon fields), which remains massless
in the color superconducting phase. Therefore, a color
superconductor would be penetrated by an external magnetic field
without restricting them to quantized flux tubes.

\section{Effects of $B$ on hypothetical combustion processes}

\subsection{Flames, instabilities, and asymmetries}

Combustion processes are common in astrophysics. A classic example
is type Ia supernovae in which a thermonuclear explosion
completely disrupts a white dwarf. Combustion processes maybe also
relevant for neutron stars. In some situations (see e.g.
\cite{Benvenuto1989,Lugones1999}), part or the whole neutron star
can convert into quark matter. This conversion is believed to
proceed as a combustion process
\cite{Benvenuto1989,Lugones1994,Lugones2002apj}. The combustion
starts as a laminar  deflagration (slow combustion) propagating
outward, and then enters a regime of turbulent deflagration  due
to the action of several hydrodynamic instabilities, such as the
Landau-Darrieus (LD) and the Rayleigh-Taylor (RT) instability, in
which the flame wrinkles and accelerates. For the largest scales
(those much larger than the thickness of the flame), the RT
instability is expected to dominate over LD in the astrophysical
case. These effects can be understood within a fractal model in
which the flame behaves like a fractal with an area $\bar A
\propto \bar{R}^{D}$ (with $2\leq D<3$), where $D$ is the fractal
dimension of the surface, and $\bar{R}$ is the mean radius of the
wrinkled surface \cite{Filyand}. Numerical simulations
\cite{Filyand} and laboratory experiments involving different gas
mixtures \cite{Gostintsev} show that the fractal growth actually
increases the velocity of the combustion front because of the
change in the transport mechanism from a laminar to a fully
turbulent burning.

The effect of the magnetic field on flames can also be understood
within a fractal model \cite{Ghezzi2001}. For the sake of
simplicity, we shall assume hereafter a dipolar magnetic field
geometry and thus the flame propagating in two particularly
representative directions, one parallel to the B-field lines (in
the polar direction) and the other one perpendicular to them (in
the equatorial direction). Within the fractal description, the
effective velocity of the flame  is given by $v_{f} = v_{lam} ( L
/l)^{D-2}$ where  L and $l$ are the maximum an minimum length
scales of perturbations unstable to the Raleigh-Taylor
instability. The characteristic velocity of the Raleigh-Taylor
growing modes must be $\ge v_{lam}$, i. e.  $l n_{RT}(l) =
v_{lam}$, where $n_{RT}$ is the inverse of the characteristic RT
time. In the absence of a B-field $n_{RT}= (1/2\pi) \sqrt{gk
\Delta \rho /2 \rho}$ and so $l = 4 \pi \rho v^2_{lam}/g \Delta
\rho$. However, in the equatorial direction, the presence of B is
essential in modifying  the dispersion relation for the RT
instability which reads $n_{RTB} =1/ 2\pi \sqrt{ gk (\Delta \rho
/2 \rho - k B^2/4 \pi g \rho ) }$ where  $k= 2 \pi /\lambda$,
$\lambda$ is the wavelength of the perturbation, and $\Delta\rho =
\rho_u - \rho_d$ is the density difference  between the upstream
and downstream parts of the flame front \cite{Ghezzi2001}. Taking
into account the velocity of the flame, the minimum cut-off length
is given by the condition $l_{e} n_{RTB}(l_{e}) = v_{lam}$, from
which we derive the minimum scale in the equatorial direction
$l_{e} = 8 \pi( B^2/8 \pi +   \rho v^2_{lam}/2  ) / g \Delta
\rho$. So, from the above definition of the fractal velocity we
have $v_{e} = v_{lam} ( L /l_{e})^{D-2}$ and $v_{p} = v_{lam} ( L
/l_{p})^{D-2}$. Note that L is not modified by B and has the same
value in both directions. Then, the ratio between the equatorial
and polar velocities is \cite{Ghezzi2001}:
\begin{equation}
\xi  = \frac{v_p }{v_e}= \bigg[ 1 + \frac{B^2 }{ 4 \pi {\rho}  \,
v_{lam}^{2}}  \bigg]^{D-2} .
\end{equation}

\noindent That is, although the magnetic pressure $B^2 /(8 \pi)$
is not relevant compared to the degeneracy pressure of neutron
star matter, the $B$-field is  important for the combustion
because it quenches the growth of RT instabilities in the
equatorial direction acting as a ``surface tension'', while it is
innocuous in the polar one where in average we have $\vec{v}_{p}
\times \vec{B} = 0$ . More specifically  $B$ modifies the minimum
RT instability scale and since the turbulent flame velocity is
related to RT-growth this results in a different velocity of
propagation along each direction. Notice that although this is
based solely on a linear analysis, 3D numerical simulations of the
development of the magnetic RT instability \cite{Jun1995}
reinforce the results of asymmetry here reported and also reveals
a tendency for its amplification in the non-linear regime. This
has interesting effects for thermonuclear supernovae
\cite{Ghezzi2001,Ghezzi2004} and neutron stars
\cite{Lugones2002apj,Lugones2005}.

\subsection{Gamma-ray bursts from neutron star phase transitions}

It has long been recognized that the transition to quark matter
inside neutron stars provides a suitable inner engine for GRBs
since some key requirements of observations are fulfilled (e.g.
energy released \cite{Bombaci2004,Ouyed2005}, timescale of the
gamma emission \cite{LugonesAPJL}, small baryon loading
\cite{Paczynski2005}, rate of events \cite{KSCheng1996}). The
total energy released is $\sim 10^{53}$ ergs \cite{Bombaci2004},
and can be even larger if the final state is paired quark matter,
due to the liberation of the gap energy. The timescale of gamma
emission in this model has been calculated to be $\sim 0.2$ s
\cite{Haensel1991,LugonesAPJL,Ouyed2005}, i.e. compatible with
short gamma ray bursts and giant flares in soft gamma repeaters
\cite{Lugones2005}. Another interesting feature is that the
surface of a quark star can guarantee an environment with a very
small baryon load, automatically generating conditions needed for
an ultra-relativistic fireball \cite{Paczynski2005}.

The magnetic field is essential for producing a collimated gamma
emission in this class of models due to the effect in the
hydrodynamics of the conversion process \cite{LugonesAPJL}. The
basics of the proposed mechanism are as follows
\cite{LugonesAPJL}. A seed of quark matter may become active or
form following the standard supernova bounce. The interface must
then propagate outwards powered by the energy release of converted
neutrons, much in the same way as a laboratory combustion. It
seems reasonable to assume the combustion to begin as a laminar
deflagration, which quickly reaches a regime of turbulent
deflagration. The magnetic field generates a strong acceleration
of the flame in the polar direction. This results in a
(transitory) strong asymmetry in the geometry of the just formed
core of hot quark matter. While it lasts, this geometrical
asymmetry gives rise to a bipolar emission of the thermal
neutrino-antineutrino pairs produced in the process of quark
matter formation. This is because almost all the thermal neutrinos
generated in the process of quark matter formation will be emitted
in a free streaming regime through the polar cap surface, and not
in other directions due to the large opacity of the matter
surrounding the cylinder. Further annihilation into
electron-positron pairs just above the polar caps, gives rise to a
relativistic fireball. Since the timescale of gamma emission is
$\sim 0.2$ s, this provides a suitable explanation for the inner
engine of short gamma ray bursts.



\begin{thebibliography}{99}

\bibitem{books} R.N. Manchester and J.Taylor, Pulsars (Freeman, 1977)

\bibitem{Lyne1998} A. G. Lyne and F. Graham-Smith, Pulsar Astronomy (Cambridge Univ. Press, 1998).

\bibitem{shapiro} S. L. Shapiro and S. A. Teukolsky, Black holes,
white dwarfs and neutron stars (Wiley, 1983);

\bibitem{mclaughlin} M. A. McLaughlin et al., ApJ 591, L135 (2003)

\bibitem{database} see  http://www.atnf.csiro.au/research/pulsar/psrcat/

\bibitem{bignami2003} G. F. Bignami, P. A. Caraveo, A. De Luca and  S.
Mereghetti, Nature 423, 725 (2003)

\bibitem{snr1}  D. Marsden, R. E. Lingenfelter and R. E. Rothschild,
 Mem. Soc. Astron. Ital. 73, 566  (2002)

\bibitem{snr2}   B. M. Gaensler and D. A. Frail, Nature 406, 158 (2000);
S. E. Thorsett, W. F. Brisken and W. M. Goss, ApJ 573, L111
(2002).

\bibitem{Link-Epstein}  B. Link,  R.I. Epstein and J.M. Lattimer,
Phys. Rev. Lett. 83, 3362 (1999)

\bibitem{Weber-Glendenning} N.K. Glendenning,  S. Pei and F. Weber,
Phys. Rev. Lett. 79, 1603 (1997).

\bibitem{AllenHorvath} M.P. Allen and  J.E. Horvath, ApJ. 488, 409
(1997).

\bibitem{Bdecay} T. M. Tauris and S. Konar, A\&A 376, 543 (2001)

\bibitem{pavlov2002} G. G. Pavlov,  V. E. Zavlin,  D. Sanwal and   J. Tr\"umper,  ApJ
 569, L95 (2002)

\bibitem{camilo} F. Camilo,  V. M. Kaspi, A. G. Lyne, R. N. Manchester,
J. F. Bell, N. D'Amico, N. P. F. McKay and F. Crawford, ApJ 541,
367 (2000)

\bibitem{baring} M. G. Baring and A. K. Harding, ApJ 507, L55 (1998),
\textit{ibid} 547, 929 (2001); M. G. Baring, Adv. Space Res. 33,
552 (2004) and references therein.

\bibitem{propeller} see e.g.  K. Y. Eksi  and M. A. Alpar, Astrophys. J. 620,
390 (2005); E. G. Blackman and R. Perna, Astrophys. J. 601, L71
(2004); K. Menou, R. Perna and L. Hernquist, Astrophys. J. 559,
1032 (2001); R. Perna, J. S. Heyl,  L. E. Hernquist, A. M. Juett
and D. Chakrabarty,  Astrophys. J. 557, 18 (2001) and references
therein.





\bibitem{Johnson1949}  M. H. Johnson, and B. A. Lippmann,  Phys. Rev. 76, 828
(1949).

\bibitem{Illarionov1980} C. Alcock and A. Illarionov,  Astrophys. J. 235, 541 (1980).

\bibitem{Lai2001} D. Lai,   Rev. Mod. Phys. 73, 629-661, (2001).

\bibitem{Zavlin2002} V.E. Zavlin and G.G. Pavlov, Proceedings of the 270 WE-Heraeus
Seminar on Neutron Stars, Pulsars, and Supernova Remnants. MPE
Report 278. Edited by W. Becker, H. Lesch, and J. Trümper.
 Max-Plank-Institut für extraterrestrische
Physik, Garching bei München, p. 263 (2002)

\bibitem{Pavlov1978} G. G. Pavlov and  Iu. A. Shibanov,  Soviet Astronomy 22,  214-222
(1978)

\bibitem{Shibanov1996} Y. A. Shibanov and D. G. Yakovlev, Astronomy and Astrophysics 309, p.171-178 (1996)

\bibitem{Potekhin2005} A. Y. Potekhin, D. Lai, G. Chabrier, and  W.C.G. Ho,
arXiv:astro-ph/0501467  (2005)


\bibitem{Ginzburg1950} V. L. Ginzburg and L.D. Landau, Zh. Eksp. Teor. Fiz. 20, 1064 (1950).

\bibitem{Abrikosov1957} A. A. Abrikosov, Sov. Phys. JETP 5,  1174
(1957); A. A. Abrikosov, Nobel Lecture, December 8 (2003)

\bibitem{Tanigawa2004} T. Tanigawa, M. Matsuzaki and  S.  Chiba, Phys. Rev. C 70, 065801
(2004)

\bibitem{Zhitnitsky2005} K. B. W. Buckley, M. A. Metlitski, and A. R.
Zhitnitsky,  Phys. Rev. C 69, 055803 (2004)

\bibitem{Link2003} B. Link, Phys. Rev. Lett. 91, 101101 (2003)

\bibitem{Lugones2002prd}  G. Lugones and J. E. Horvath,  Phys. Rev.
D66, 074017 (2002)

\bibitem{Lugones2003aa} G. Lugones  and  J. E. Horvath, Astronomy and Astrophysics,
403, 173 (2003)

\bibitem{Bailin1984} D. Bailin and A. Love, Phys. Rep. 107, 325 (1984).

\bibitem{newgap} M. G. Alford, K. Rajagopal and F. Wilczek, Phys. Lett. B
422, 247 (1998); R. Rapp, T. Sch\"afer, E. V. Shuryak and M.
Velkovsky, Phys. Rev. Lett. 81, 53 (1998); M. G. Alford, Ann. Rev.
Nucl. Part. Sci. 51, 131 (2001).


\bibitem{Iida2002} K. Iida and G. Baym, Phys. Rev. D 66, 014015 (2002); Erratum:
Phys. Rev. D 66, 059903 (2002).

\bibitem{Blashke2002}  D. M. Sedrakian and D. Blaschke,
Astrofiz. 45, 203 (2002); e-Print Archive: hep-ph/0205107 (2002).

\bibitem{Alford2000} M. Alford, J. Berges and K. Rajagopal, Nucl. Phys. B 571, 269
(2000).

\bibitem{Giannakis2003} I. Giannakis and H.-C. Ren, Nucl. Phys. B669, 462 (2003).

\bibitem{Iida2004} K. Iida, e-Print Archive: hep-ph/0412426 (2004).


\bibitem{Ruderman2004} M. Ruderman, arXiv:astro-ph/0410607 (2004)

\bibitem{Alford1999} M. Alford, K. Rajagopal and F. Wilczek, Nucl. Phys. B537, 443
(1999)

\bibitem{Benvenuto1989} O. G. Benvenuto and J. E. Horvath,   Phys. Rev. Lett. 63, 716
(1989)

\bibitem{HB}  J. E. Horvath and O. G. Benvenuto, {\it Phys. Lett. B} {\bf 213}, 156 (1988)

\bibitem{Lugones1994} G. Lugones, O. G.Benvenuto and H. Vucetich, {\it Phys. Rev. D} {\bf 50},
6100 (1994)

\bibitem{Lugones2002apj} G. Lugones, C. R. Ghezzi, E. M. de Gouveia Dal Pino and J.E. Horvath,
{\it Astrophys. J.}{\bf 581}, L101 (2002)

\bibitem{Lugones1999} O. G. Benvenuto  and  G. Lugones, {\it Mon. Not. R.A.S.}{\bf 304}, L25 (1999)

\bibitem{Filyand}  L. Filyand,  G. I. Sivashinsky and  M. L. Frankel,   Physica D 72, 110
(1994)

\bibitem{Gostintsev} Yu. A. Gostintsev,  A. G. Istratov and Yu. V. Shulenin,  Combustion
Explosions Shock Waves 24, 70 (1988)


\bibitem{Ghezzi2001}  C. R. Ghezzi, E. M. de Gouveia Dal Pino and  J. E. Horvath,  ApJ 548,
L193 (2001)

\bibitem{Ghezzi2004}  C. R. Ghezzi, E. M. de Gouveia Dal Pino and  J. E. Horvath,
{\it Mon. Not. R.A.S.} ~ {\bf 348}, 451 (2004)

\bibitem{Jun1995}  B. I. Jun, M. L. Norman, and J. M. Stone,  ApJ 453, 332
(1995)

\bibitem{Prakash} A.E. Broderick, M. Prakash and J.M. Lattimer,  Phys. Lett. B 531, 167 (2002)


\bibitem{Bombaci2004}  I. Bombaci, I. Parenti, I. Vida\~na, Astrophys.J. 614,
314 (2004)

\bibitem{Haensel1991}  P. Haensel ,   B. Paczynski  and P. Amsterdamski, {Astrophys. J.} {375}, 209 (1991)

\bibitem{KSCheng1996} K. S. Cheng and Z. G. Dai, Phys. Rev. Lett. 77, 1210 (1996)

\bibitem{Paczynski2005}     B. Paczynski and P. Haensel, arXiv:astro-ph/0502297 (2005)

\bibitem{Ouyed2005} R. Ouyed, R. Rapp, and C. Vogt, arXiv:astro-ph/0503357 (2005)

\bibitem{LugonesAPJL} G. Lugones, C.R. Ghezzi, E.M. de Gouveia Dal Pino and J.E.
Horvath, {\it Astrophys. J.}{\bf 581}, L101 (2002)

\bibitem{Lugones2005} G. Lugones, J.E. Horvath and E.M. de Gouveia Dal Pino (2005) in preparation.



\end{thebibliography}
\end{document}